
\input jnl.tex
\input reforder.tex
\def\lrg{SU(2)_L\times SU(2)_R \times U(1)_{B-L}}
\def\fp{f_{\phi}}
\def\fc{f_{\chi}}
\def\pdp{\phi^{\dag} \phi}
\def\cdc{\chi^{\dag} \chi}
\def\lam{\lambda}
\rightline{\rm Fermilab-Pub-92/228-A}
\rightline{\rm HUTP-92/A032}
\rightline{\rm hep-ph/9208245}
\title
{Metastable Cosmic Strings in Realistic Models}
\vglue 0.25 truein
\author
{Richard Holman}
\affil
{Department of Physics, Carnegie Mellon University, Pittsburgh PA 15213.}
\smallskip
\author
{Stephen Hsu\footnote{*}{Junior Fellow, Harvard Society of Fellows}}
\affil
{Lyman Laboratory of Physics, Harvard University, Cambridge, MA 02138}
\smallskip
\author
{Tanmay Vachaspati}
\affil
{Tufts Institute of Cosmology, Department of Physics and Astronomy,
Tufts University, Medford, MA 02155.}
\smallskip
\author
{Richard Watkins\footnote{**}{Address after Sept. 1: Physics Department,
University of Michigan, Ann Arbor MI 48109}}
\affil
{Department of Astronomy and Astrophysics, Enrico Fermi Institute,
The University of Chicago, Chicago, IL 60637
and
NASA/Fermilab Astrophysics Center, Fermi National Accelerator Laboratory,
Batavia, IL 60510}
\abstract
\doublespace
We investigate the stability of the electroweak Z-string at high temperatures.
Our results show that while finite temperature corrections can improve the
stability of the Z-string, their effect is not strong enough to stabilize the
Z-string in the standard electroweak model. Consequently, the Z-string will be
unstable even under the conditions present during the electroweak phase
transition. We then consider phenomenologically viable models based on the
gauge group $SU(2)_L \times SU(2)_R \times U(1)_{B-L}$ and show that metastable
strings exist and are stable to small perturbations for a large region of the
parameter space for these models.  We also show that these strings are
superconducting with bosonic charge carriers. The string superconductivity may
be able to stabilize segments and loops against dynamical contraction. Possible
implications of these strings for cosmology are discussed.

\endtopmatter

Over the last two decades, cosmic strings have evoked a great deal of interest
both as possible remnants of a Grand Unified era in the early universe as well
as a possible mechanism for structure formation in the universe\refto{av}.
However, no compelling particle physics models exist that give rise to such
defects. Recently, a defect that is closely related to the ``ordinary'' cosmic
string has been found\refto{tv91, ynetc} in what may be the most compelling of
all particle physics models - the standard electroweak model. The defect is
identical in its structure to the cosmic string solution found by Nielsen and
Olesen\refto{hnpo} and may be thought of as a cosmic string embedded in the
standard electroweak model\refto{tvmb}. The difference now is that the defect
does not owe its existence to topology and consequently may not be stable. The
stability of the defect depends on the parameters of the electroweak
model\refto{com1}.

In Ref. \cite{mjlptv} the stability of the string in the standard
electroweak model was analyzed. This resulted in a map of parameter
space demarcating the regions in which the string is stable to small
perturbations from
the regions in which it is unstable. Given the known value of the
Weinberg angle, $\sin^2 \theta_W \approx 0.23$, and the constraints on
the Higgs mass, $m_H > 57$ GeV, it is clear that the
electroweak string is unstable. However, the analysis in Ref. \cite{mjlptv}
was limited to the bare electroweak model. The issue of stability must
be reconsidered when one takes quantum and thermal corrections to
the potential into account. In essence, the question is whether
the strings can be stable at
temperatures close to the electroweak phase transition temperature.
If this is true, the strings may be relevant to cosmology. We
answer this question in Sec. 2 where we map the parameter space as in
Ref. \cite{mjlptv} for the case when thermal and quantum corrections
are taken into account.  The results
show that these corrections increase the region of stability, but not
to the extent of allowing for stable electroweak strings in the standard model
even near the electroweak phase transition.

We then consider the question of whether there is any realistic
particle physics model in which one might expect stable embedded
strings. We show that left-right models\refto{mohapatra} are good
candidates. In Sec. 3 we consider an $SU(2)_L \times SU(2)_R \times
U(1)_{B-L}$ model. The parameters may be chosen such that this model
gives acceptable particle physics and also contains {\it stable}
strings. This gives us a
concrete example of a realistic particle physics model with stable
embedded strings.

The cosmology of embedded strings will be very different from that
of topological strings. The basic reason for this has to do with the
metastability of the embedded strings versus the complete stability
of topological strings. In Sec. 4 we speculate on the cosmology of
embedded strings. The results of this section should not be thought of
as firm conclusions but only as first guesses intended to inspire
future work. Section 5 contains our conclusions.

\beginsection{2. Stability of the Z-String at Finite Temperature}

The addition of quantum corrections to the Higgs potential can
have a drastic effect on the vacuum structure of the standard
model\refto{sher}.  The most important correction is in the form of a
$\phi^4 \log(\phi/M)$ term, which can destabilize the potential at large
$\phi$.  However, our interests lie at relatively small $\phi$
where this term is quite small.  We have done a stability
analysis including this term and found that there is very little
effect.  Therefore for the remainder of this discussion we shall
ignore quantum corrections and concentrate on those induced by
finite temperature effects.

The one loop finite temperature effective potential for the Higgs field
can be written as\refto{dlhll}
$$
V_T (\phi) = \lambda\left( \phi^{\dagger}\phi- {\eta^2\over2}\right)^2
+ DT^2\phi^{\dagger}\phi - ET(\phi^{\dagger}\phi)^{3/2} \eqno(2.1)
$$
where $D$ and $E$ are functions of the particle masses, and can
be approximated by
$$\eqalignno{
D&= {1\over 8\eta^2}\left( 2m_W^2+ m_Z^2 + 2m_t^2\right), &(2.2)\cr
E&= {1\over 4\pi \eta^3}\left( 2m^3_W + m_Z^3\right) &(2.3)\cr .}
$$
Here $\eta=246$ GeV is the expectation value of ${\phi}$ at the minimum
of the zero-temperature potential. Here we have chosen to ignore temperature
corrections to $\lambda$, which are logarithmic and should not
effect our results significantly.

As in the zero temperature case, Z-string solutions will take
the form
$$
W^{\mu 1}=W^{\mu 2}= A^{\mu}=0,\qquad Z^{\mu}= -{v(r)\over r}\hat
e_{\theta}
$$
$$
\phi = f(r)e^{i\theta} \Phi ,\qquad \Phi = {0\choose 1},
\eqno (2.4)
$$
where we have assumed the string to be aligned along the $z$
axis, and $r$ and $\theta$ are polar coordinates in the
$xy$-plane. The functions $f$ and $v$ are determined by the equations of
motion:
$$
f^{\prime\prime} + {f^{\prime}\over r} - \left( 1- {\alpha\over 2}v\right)^2
{f\over r^2} - 2\lambda\left(f^2 - {\eta^2\over 2}\right)f
+ DT^2f - {3\over 2}ETf^2=0
\eqno(2.5)
$$
$$
v^{\prime\prime}- {v^{\prime}\over r} +
\alpha\left(1-{\alpha\over 2}v\right)f^2=0
\eqno(2.6)
$$
where primes denote differentiation by $r$.  Here
$\alpha$ is given by $g= \alpha \cos(\theta_W)$.

The functions also
satisfy the boundary conditions:
$$
f(0)=v(0)=0,\qquad f(\infty)= f_{\infty},\qquad v(\infty)= {2\over \alpha}
\eqno (2.7)
$$
where $f_{\infty}$ is the magnitude of the global minimum of $V_T$.


In order to study the solutions to these equations numerically, it
is convenient to introduce the dimensionless quantities:
$$
P\equiv f/f_{\infty} ,\qquad V\equiv {\alpha\over 2}v,\qquad R\equiv
{\alpha\eta
\over 2\sqrt2}r
\eqno(2.8)
$$
so that the equations take the simple form
$$
P^{\prime\prime} + {P^{\prime}\over R} - \left( 1- V\right)^2
{P\over R^2} - \beta(P-1)(P-P_e)P = 0.
\eqno(2.9)
$$
$$
V^{\prime\prime}- {V^{\prime}\over R} +
\left(1-V\right)P^2=0
\eqno(2.10)
$$
where $\beta = {8\lambda\over \alpha^2}$ and
$$
P_e= -1 + {2ET\over
ET+ \sqrt{T^2\left( E^2- {4\beta\alpha^2\over9}D\right)
+ {\beta^2\alpha^4\over 18}\eta^2}.}
\eqno(2.11)
$$

The parameter $P_e$ carries all of the information about finite
temperature effects in the rescaled potential.  It takes on values
between $-1$ at $T=0$ up to $0.5$, above which $P=0$ becomes
the true vacuum.  For $0 < P_e < 0.5$, $P=0$ is a local minimum,
separated from the global minimum at $P=1$ by a potential
barrier.  This assumes the phase transition to be first order;
in models with a second order transition $P_e \le 0$.

Eqns. (2.9-2.10) can be solved using standard methods.  The string
configurations that result, even in the extreme $P_e=0.5$ case,
are not qualitatively different than $T=0$ strings.  In
particular, both $P(r)$ and $V(r)$ remain monotonically
increasing functions of $r$.

The stability of electroweak strings at finite temperature can
be determined in a similar manner to that for zero temperature
strings as described in Ref. \cite{mjlptv}.
Here we will give a short review
of the procedure, referring the reader to Ref. \cite{mjlptv}
for details.


The energy functional for two-dimensional
static solution in the electroweak model
may be written in the
standard notation of Ref. \cite{jct}:
$$
E = \int d^2 x \left [
          \fourth G_{ij} ^a G_{ij} ^a + \fourth F_{Bij} F_{Bij}
          + (D_j \phi ) ^{\dag} (D_j \phi ) + V_T (\phi )
              \right ]
\eqno (2.12)
$$
where, $i,j = 1,2$ and $a = 1,2,3$.
The string solution that extremizes this energy
functional is given in (2.4) and we now perturb around that solution.

It can be shown that the only relevant perturbations are those in
which the upper component, $\phi_1$,
of the Higgs doublet and the $W$ fields
are perturbed. These fields can be expanded in modes:
$$
\phi _1 = \chi^m (r) e^{im\theta}
\eqno (2.13)
$$
for the $m^{th}$ mode where $m$ is any integer. For the
$n^{th}$ mode of the gauge fields
we have,
$$
{\vec W}^1 = \left [
\left \{ {\bar f}^n_{1} (r) \cos(n\theta ) + f^n_{1} \sin(n\theta ) \right \}
{\hat e} _r +
{1 \over r} \left \{
           - {\bar h}^n_{1} \sin(n\theta ) + h_{1}^n \cos(n \theta ) \right \}
{\hat e} _\theta
\right ]
\eqno (2.14)
$$
$$
{\vec W}^2 = \left [
\left \{ - {\bar f}^n_{2} (r) \sin(n\theta ) + f_{2}^n \cos(n\theta ) \right \}
{\hat e} _r +
{1 \over r} \left \{
           {\bar h}_{2}^n \cos(n\theta ) + h_{2}^n \sin(n \theta ) \right \}
{\hat e} _\theta
\right ]
\eqno (2.15)
$$
Of these, the $m=0, \ n=1$ mode is the most dangerous and it is
sufficient to look at this mode alone. Further, it turns out that the
barred variables separate from the unbarred ones and it is sufficient
to look at the problem in the unbarred variables alone.

Define the quantities
$$
F_{\pm} = {{f_2 \pm f_1} \over {2}}
\eqno (2.16)
$$
$$
\xi_{\pm} = {{h_2 \pm h_1} \over 2}
\eqno (2.17)
$$
$$
\zeta = (1-\gamma v) \chi + \half g f \xi_+
\eqno (2.18)
$$
Then, after substituting the mode expansions in the energy functional
and a lot of algebra, we find that the energy variation around the
unperturbed solution is,
$$
\eqalign{
\delta E = 2\pi \int dr r \biggl [ \biggl \{
  {{\zeta '}^2 \over {P_+}} + & U(r) \zeta ^2 \biggr \} +
\cr
&
{\rm sum \ of \  whole \  squares} \biggr ]
\cr
}
\eqno (2.19)
$$
where primes denote differentiation with respect to $r$,
$$
P_+ = (1 - \gamma v )^2 + \half g^2 r^2 f^2 \  ,
\eqno (2.20)
$$
$$
U(r) = {{{f'}^2} \over {P_+ f^2}} + {{2 S_+} \over {g^2 r^2 f^2}} +
          {1 \over r} {d \over {dr}} \biggl (
                     {{r f'} \over {P_+ f}} \biggr ) \  ,
\eqno (2.21)
$$
and,
$$
S_+ = {{g^2 f^2} \over 2} - {{\gamma ^2 {v'}^2} \over {P_+}} +
      \gamma r {d \over {dr}} \left [
                  {{v'} \over r} {{(1-\gamma v)} \over {P_+}}
                             \right ] \   .
\eqno (2.22)
$$
The energy variation in (2.19) is minimized if the sum of squares
is chosen to vanish. This fixes the modes
$F_\pm$ and also requires $\xi_- = 0$. Then we are simply left
with a problem in $\zeta$. (Note that although the explicit form
of the Higgs potential does not appear anywhere in the $\zeta$
dependent terms, it does appear implicitly in the unperturbed
solution given by $P$ and $V$.)

By performing an integration by parts, we
can now rewrite the $\zeta$ part of $\delta E$ as,
$$
\delta E[ \zeta ] = 2\pi \int dr \  r
              \zeta {\bf O} \zeta
\eqno (2.23)
$$
where, {\bf O} is the differential operator:
$$
{\bf O} = - {1 \over r} {d \over {dr}} \left (
             {r \over {P_+}} {d \over {dr}} \right ) + U(r) \  .
\eqno (2.24)
$$
Now we have to determine if the differential equation
$$
{\bf O}\zeta = \omega \zeta
\eqno (2.25)
$$
has any negative eigenvalues ($\omega$). The boundary conditions
on $\zeta$ are:
$\zeta (r=0) = 1$ and $\zeta (r = \infty ) = 0$.

In this way, the whole
stability analysis has been reduced to the single differential
equation (2.25). This equation can be solved numerically using
the shooting method.



For a given $P_e$ and $\beta$, one can use (2.25) to find the
value of $\theta_W$ for which there exists a zero
eigenmode.  This determines the critical value of $\theta_W$
which marks the boundary between stability and
instability for the strings.  In Fig. 1 we plot a number of
curves showing the regions of stability for different values of
$P_e$.  In Fig. 2, we plot regions of stability for different
temperatures, assuming fixed values of $D/\alpha^2$ and
$E/\alpha^2$.


These results show a number of significant features.
First, we see that the region of stability grows significantly
as the temperature increases, including regions where $\beta>1$.
Thus finite temperature effects
have a stabilizing effect on the strings.  This allows one to
consider scenarios where strings only exist during a specific
epoch in the evolution of the universe.  Stable strings can form in a phase
transition, but then become unstable and disappear as the
temperature drops below some critical value.  It is important to
note, however, that even at finite temperature the standard
model value of $\sin^2(\theta_W)=0.23$ and $m_H > 57 GeV$
is still deep within the
region of instability.


Another feature of the plot is that there appears to be a lower
bound to the $\sin^2(\theta_W)$ at which there are stable
strings.  The stability region for small $\beta$ is unknown, due
to the fact that in this limit the strings become very thick and
are difficult to treat numerically.  For the $T=0$ case it was
unclear whether the region of stability extended down to small
$sin^2(\theta_W)$ at small $\beta$.  For the finite temperature
case, all of the stability curves corresponding to different
$P_e$ should converge at $\beta=0$.  This is because the
potential becomes unimportant in the lagrangian
in this limit.  From the figure it seems
likely that the critical value of $sin^2(\theta_W)$ at $\beta=0$
is around $0.92$.  The absolute lower bound for stable strings
is thus given by the $P_e=0.5$ curve, and is $\simeq 0.91$.

\beginsection{3. Metastable Strings in Left-Right Models}

As we have seen in Sec. 2, the Z-strings of the electroweak theory are
unstable to small perturbations for physically reasonable values of the Higgs
mass and the Weinberg angle, even when quantum
and finite temperature corrections
are taken into account. This leads us to wonder if there are {\it any}
well motivated particle physics models that admit (meta) stable, embedded
strings.

A hint as to how to go about finding such a model comes from the analysis in
Sec.2.  There we were essentially trying to stabilize the string
through modification of the Higgs potential. Our failure to
obtain stable strings was due in some part to the small value of
$\sin^2(\theta_W)$ required by the standard model.  Thus, we are motivated
to look for extensions of the standard model where the gauge
sector of the theory is enlarged, allowing for the presence
of other Weinberg-type angles which can take somewhat larger values.


A well-known extension of the gauge sector of the standard model is the
left-right model based on the gauge group $\lrg$\refto{mohapatra}. One of the
interesting features of this class of theories is that they can be compatible
with known experimental results even if the scale of $SU(2)_R\times U(1)_{B-L}$
breaking to $U(1)_Y$ is quite low i.e. $500$ GeV - $1$ TeV.\refto{buchmuller}

The field content of the model we consider is as follows (the quantum numbers
are the $\lrg$ representation assignments). The left handed quarks and leptons
transform as $(2, 1; 1/3),\ (2, 1; -1)$ respectively while their right-handed
companions transform as $(1, 2; 1/3),\ (1, 2; -1)$ (note that we have added a
right handed neutrino state). The minimal Higgs content required for the
phenomenological viability of the model is:
$\phi \sim (1, 2;-1),\ \chi \sim (1, 3; 2),\ \Delta \sim (2, 2; 0)$. The right
handed triplet $\chi$ is required to give the right handed neutrino a large
Majorana mass so as to implement the see-saw mechanism\refto{seesaw}, while
$\phi$ is needed to yield the correct pattern of symmetry breaking and
$\Delta$ induces the Dirac masses of all other fermions. The phenomenology of
this model was considered in Ref.\cite{buchmuller}.
We assume that the
vacuum expectation values (VEV's) $\fp$, $\fc$, $v$ of $\phi$, $\chi$ and
$\Delta$ respectively satisfy the following hierarchy: $v<<\fc<<\fp$.

 From the viewpoint of constructing models with embedded metastable strings,
the advantages of the left-right model described above are clear. We can just
take the Z-string found in the electroweak model and embed it in the
{\it right handed} sector. To the extent that $v,\ \fc$ are much smaller that
$\fp$, we can neglect the backreaction of $\Delta$ and $\chi$ on the string
configuration described by $\phi$ and the right-handed neutral gauge boson
$Z_R$ (these effects are proportional to $\fc/\fp$). Thus the stability
analysis
of Sec.2 goes through without any changes
except for the replacement $(g, g')\rightarrow (g_R, g_{B-L})$. This implies
that there is a non-trivial region of the parameter space for which our model
will admit metastable strings.

While the strings of the left-right model are stable to small perturbations, as
it stands, it would appear that they are unstable to perturbations along the
string.
In other words, they are unstable to contraction (which leads to the
annihilation of the monopole-antimonopole pair at the ends of the string).
Since we expect that the contraction time scale will be at most a Hubble time,
if these strings are to be of any cosmological significance, we must find a way
to make them more stable against this mode of instability. One
possibility is that if the strings are {\it superconducting},
a standing wave
of charge carriers can be set up along the string, which reflects off the
monopoles at either end. While the reflection coefficient is not unity, we can
imagine that it is large enough so that it will take  some time before the
string can rid itself of enough current so as to allow for it to contract away.
This mechanism for preventing the dynamical collapse of string {\it loops}
has been studied in some detail in earlier work and it has been shown that
there is a region of parameter space where current carrying loops can
form static rings, or, vortons. As we
will see below, it is these loops that will be most relevant from a
cosmological perspective.

To show that the string is superconducting we start by displaying the Higgs
potential for the coupled $\phi-\chi$ system (note that this system is
remarkably similar to the triplet majoron model\refto{gelminironc, georgi}):
$$
\eqalign{V(\phi, \chi)&=\lam_1 (\pdp - \fp^2/2)^2 +
\lam_2 (tr(\cdc) - \fc^2/2)^2 +
\lam_3 (\pdp - \fp^2/2) (tr(\cdc) - \fc^2/2)\cr
+& \lam_4 (\pdp tr(\cdc) - \phi^{\dag}\chi \chi^{\dag} \phi)+
\lam_5 ((tr(\cdc))^2 - tr(\cdc \cdc))}.
\eqno(3.1)
$$
We parametrize $\phi$ and $\chi$ in the following fashion:
$$
\phi = \pmatrix{\phi^{0}\cr
                \phi^{-}\cr},
\eqno(3.2)
$$
$$
\chi = \pmatrix{\chi^{+}/\sqrt{2}&
                \chi^{++}\cr
                \chi^{0}&-\chi^{+}/\sqrt{2}\cr},
\eqno(3.3)
$$
where the factors of $\sqrt{2}$ in $\chi$ have been chosen so that if
$D_{\mu} \chi$ is the $\chi$ covariant derivative, then
$tr((D_{\mu}\chi)^{\dag} D^{\mu}\chi)$ is the correctly normalized $\chi$
kinetic term. Note that $\chi$ can also be written as
$\chi = \sqrt{2}(T_3 \chi^{+} + T_{+}\chi^{++} + T_{-}\chi^{0})$, where
$\{T_{+}, T_{-}, T_3\}$ are the generators of $SU(2)_R$, satisfying;
$[T_3, T_{\pm}] = \pm T_{\pm}, \ [T_{+}, T_{-}] = T_3$. The action of $SU(2)_R$
on $\chi$ is via commutator:
$T^{a} W^{a}_{\mu}\cdot \chi \equiv [T^{a}, \chi]W^{a}_{\mu}$

If $\lam_{1,2,4,5} >0$ and $|\lam_3|<2 \sqrt{\lam_1 \lam_2}$, then
$V(\phi, \chi)$ is positive definite and $\phi$, $\chi$ acquire the following
VEV's: $\langle \phi^{0} \rangle = \fp/\sqrt{2},\  \langle \chi^{0} \rangle =
\fc/\sqrt{2}$.

We now claim that given this potential, there are large regions of the
$\lam_{1, 2, 3, 4, 5}$ parameter space for which $\chi^{+, ++}$ act as
bosonic
charge carriers on the $\phi$ string. To show this, we use the following
argument, first given by Witten\refto{witten}. First we show that it can be
energetically favorable for the components of $\chi$ to be nonzero in the core
of the string where $\phi = 0$. If $\phi = 0$, the potential for $\chi$ reads:
$$
 \eqalign{V(\phi=0, \chi)& = -\lam_1 \fp^2/2 + \lam_2 (tr(\cdc) - \fc^2/2)^2
\cr -&
\lam_3 (\fp^2/2) (tr(\cdc) - \fc^2/2) +
\lam_5 ((tr(\cdc))^2 - tr(\cdc \cdc))}
\eqno(3.4).
$$
This is extremized if (i) $\chi = 0$, or (ii) either $|\chi^{0}|^2 =
|\chi^{++}|^2$ or $2\lam_2 (tr(\cdc)-\fc^2/2) = \lam_2 \fp^2/2$. It is easy to
see that if $\lam_2 \fc^2 + \lam_3 \fp^2/2>0$, then $\chi = 0$ is a {\it
maximum} of the potential. Thus, in this case,
non-zero values of $\chi$ are energetically preferred in the string
core.

The above analysis is not sufficient to show the existence of bosonic charge
carriers. We must check to see that the kinetic term for $\chi$ also allows
for a nonzero value of $\chi$ in the string. We do this by showing that the
equations of motion for $\chi$, linearized around $\chi=0$, admit growing
solutions. This will then show that in the background of the $\phi$ string,
$\chi$ is unstable to the formation of a nonzero condensate on the string.
Let us first consider the $\chi^{++}$ equation of motion:
$$
\eqalign{
-\partial_{\mu}(\partial^{\mu}\chi^{++} -
i\alpha_R \cos 2\theta_R Z_R^{\mu}
\chi^{++})& = 2 \lam_2 (tr(\cdc) - \fc^2/2) \chi^{++}
\cr
+&
\lam_3 (f_{NO}(\rho)^2 - \fp^2/2)\chi^{++} +
\lam_5 ({\chi^{+}}^2 + 2 \chi^{0} \chi^{++}){\chi^{0}}^{*}.
\cr
}
\eqno(3.5)
$$
Here $\alpha_R \equiv \sqrt{g_R^2 + g_{B-L}^2}$, $\theta_R$ is the right-handed
version of the Weinberg angle and
$f_{NO} (\rho)$ is the $\phi$ part of the string
configuration. In the string, $Z_R$ takes the form
$Z_R(\rho) = -\left( v(\rho)/\rho\right) \hat{e}_{\theta}$, where $v(\rho)$
is the Nielsen-Olesen
configuration for the vector field. We now linearize eqn.(3.5) around
$\chi=0$ and take the following form for the perturbation of $\chi^{++}$:
$\delta \chi^{++} = \exp(-i \omega_{++} t) g_{++}(\rho)$.
The linearized equation of motion for $g_{++}(\rho)$ reads:
$$
-\nabla^2 g_{++}(\rho) + {\cal V}(\rho) g_{++}(\rho) = \omega^2_{++}
g_{++}(\rho)
\eqno(3.6),
$$
where ${\cal V}(\rho)$ is given by:
$$
{\cal V}(\rho) = -\lam_2 \fc^2 + \lam_3 (f_{\rm{NO}}(\rho)^2 - \fp^2/2)
\eqno(3.7),
$$
and $\nabla^2$ is the two dimensional Laplacian.
We see that at
$\rho = 0$, ${\cal V} = -(\lam_2 \fc^2 + \lam_3 \fp^2/2)$ and that ${\cal V}$
increases monotonically with $\rho$ until it reaches the asymptotic value of
$-\lam_2 \fc^2$. Thus, as long as $\lam_2 \fc^2 + \lam_3 \fp^2/2>0$, ${\cal V}$
is negative definite and, as in Witten's original analysis\refto{witten}, the
two dimensional Schroedinger
equation above for $g_{++}$ will admit at least one bound state with
$\omega_{++}^2<0$. Thus, $\chi^{++}$ is unstable to forming a condensate on the
string. A similar analysis can be repeated for the other components of $\chi$,
with the result that under certain conditions, they too can condense onto the
string (except for $\chi^0$, since it has a nonzero expectation value away
from the string).

\beginsection{4. Cosmological Speculations}

Here we speculate as to the possible cosmological implications of
embedded strings.
There are many uncertainties in outlining any cosmological scenario
involving these
strings because of the model dependence of many of their
characteristics.
Here we will
content ourselves with outlining one of several possible scenarios
in which embedded strings might be
cosmologically relevant. We will also try to compare and contrast
the formation and
evolution of embedded strings with that of standard topological
strings.

Consider the production of strings in a phase transition in the
early universe. At temperatures above the phase transition
temperature, the thermal fluctuations in the fields will
spontaneously produce string-like configurations which will, however,
decay just as fast. As the temperature decreases and the
universe goes through the phase transition, some of the string-like
configurations that were undergoing thermal fluctuations will freeze
out and thus not decay. It is these string configurations that may
survive the phase transition and be important for cosmology.
This process of thermal production is the same for topological
as well as embedded strings.

There is, however, an important difference between topological
and embedded strings. This is that topological strings cannot
end whereas embedded strings may end on monopoles. Hence, after
the phase transition, topological strings can only occur as
closed loops or infinite strings, while embedded strings can also
occur
as finite segments of strings with monopoles attached at their
ends. This is the crucial difference between the two kinds of
strings.

We now discuss the formation of embedded strings. The first question
is: what is the size distribution of the embedded strings
after the phase transition?
This question cannot be answered with any certainty but some
reasonable
guesses can be made. As we discussed above, the production of the
strings is thermal and is similar in some ways to the production of
topological
strings; hence, it is prudent to first look at topological strings.
In this case\refto{tvavformation},
the string network upon formation consists of
a network of infinite strings that contain about 80\% of the
entire string length. The remaining 20\% goes into a scale
invariant distribution of closed loops. These results were obtained
by using an argument first given by Kibble\refto{kibble}. In this
argument, if the boundary conditions on a spatial contour
are fixed, they determine whether there is a
string passing through the contour almost
unambiguously.\refto{com4}
So to detect the presence of a string, all one
needs to check are the boundary conditions.

This ``Kibble'' mechanism does not
apply in the case of embedded strings because the boundary conditions
are not sufficient to determine the presence of a string. However,
one might assume that there is a certain probability of a
string passing through any given contour. On this basis, one
could attempt to use the results of Ref. \cite{tvformation}.
In the case that
the probability of string formation is sufficiently
low, there is a population of loops whose length distribution
is given by,
$$
dn(l) = a {{e^{-b l/\xi }} \over {\xi^2 l^2}} dl \ .
\eqno (4.1)
$$
The dimensionless
parameters $a$ and $b$ will depend on the probability of string
formation while $\xi$ is the correlation length at the phase
transition which we will assume is given roughly by $T_c ^{-1}$
where $T_c$ is the temperature at which the phase transition
occurs. If the phase transition is second order, the
correlation length can actually exceed $T_c^{-1}$ by orders of
magnitude, leading to a
small value of $b$ in the above equation.
Note that the exponent of $l$ in the denominator is
2 and not $5/2$ as might be expected from a scale invariant
distribution. When the string formation probability is low,
the number density of open strings will be similar to that in (4.1)
but the overall amplitude will be suppressed by a factor
$\exp(- (m - \mu m^{-1}) / T_c )$ where, $m$ is the mass of the
monopole necessary
to terminate a string and $\mu$ is the mass density of the string.
(The exponent is derived by the following considerations: the energy
cost in terminating
a string is the mass of the monopole $m$ but were the string not to
terminate, the energy cost would have been the string density $\mu$
multiplied by the size of the monopole $\approx m^{-1}$. )
If the mass of the monopole
is large - that is, if the strings are stabilized by a large potential
barrier - the open segments of string are negligible in number as
compared to the closed loops. It should also be remarked that the
exponential suppression of long loops (and open segments) may
be viewed as a Boltzmann factor in the thermal production of
embedded strings.

If the probability of string formation is large, the loop
distribution
will be given by a scale-invariant distribution
$$
dn(l) = \alpha {{e^{-\beta l/\xi }} \over
{\xi ^{3/2} l^{5/2}} } dl \  .
\eqno (4.2)
$$
If strings cannot terminate, a network of infinite strings would also
be present. However, since embedded strings can terminate, the length
that
would have been in infinite strings would now be in finite segments
of string. The length distribution of the finite segments would also
be exponential since, at every step, there is a certain probability
for the string to end. But, for large string formation probabilities,
the total length in open segments would exceed the
total length in loops. In the limit that the monopole becomes
infinitely
heavy, the open segments would be infinitely long and the fraction
of length in open strings would approach 80\%.

There is another important feature of the string network that we
have ignored so far: the strings are superconducting. Then, during
the phase transition, random currents will be induced on the strings.
The net current on a loop of size $l$ is expected to be proportional
to $T_c ^2 (l/\xi_i )^{1/2}$ where $\xi_i \sim T_c ^{-1}$
is the correlation length of the
random currents. There will be currents on the open strings also.
However, it is not clear what happens to the current when it
encounters the monopole at the end of the string. We expect that
the current could be reflected off the monopole and, in this way, a
standing wave would be set up on the open string.
(In addition to the reflection, there might be a small
transmission amplitude and the current would slowly leak out
from the string.) Another way of
saying this is that the zero modes are a solution to the Dirac
equation (or the Klein-Gordon equation for bosonic superconductivity)
in the presence of the string. The string provides a potential well
for the zero mode carriers.
In the case of an
open string, one might envisage the presence of standing wave
solutions while in the case of a loop, one can imagine traveling
waves going around the loop in addition to the standing waves.

In the following we will assume that, after the phase transition,
there is a loop distribution given by (4.1) and a
strongly suppressed open string
distribution also given by the form in (4.1).
In addition, all the strings carry
currents in proportion to the square root of their length.

What is the evolution of this system?

Let us first consider the loops. The dynamics of the loop is
governed by the tension of the string, frictional forces and
the Hubble expansion. We are justified in ignoring the Hubble
expansion since it
is unimportant on scales much smaller than the
horizon. (The embedded string distribution is exponentially
suppressed at long lengths and so it is unlikely to find strings
that span the horizon. Therefore, the Hubble expansion has
no significant effect on the dynamics of embedded strings).
Initially the frictional forces are very
large and so the string motion is highly damped. This would lead
to a collapse of the loops under their own tension. However,
the effective tension of the string is a sum of the bare tension and
the
square of the current on the string\refto{ecmhnt, rdps}. As the loop
collapses, the current builds up and the effective tension becomes
smaller. Now two
possibilities can occur\refto{ecmhnt, rdps, chhhmt, abtpds}:
(i) the current is so large
that the charge carriers can leave the string - that is, the current
can saturate, and, (ii) the effective tension goes to zero and the loop
does not collapse any further. If possibility (i) is realized, the
loops continue to collapse and eventually disappear. Depending
on the lifetime of the loops and their decay products, their
cosmology may be of some interest.
If possibility (ii)
is realized, the loops form static ring configurations that can
survive until some quantum tunneling event causes the charge to
leak. In this case, the rings would have a magnetic dipole moment
and perhaps some net electric charge and could survive for a
very long time. Depending on the net charge that a ring carries,
the rather severe
constraints on charged dark matter (CHAMPS) would
apply\refto{champs}.

The evolution of the open segments of strings is even less certain
than that of the loops but we shall indicate some possible scenarios.
The initial dynamics of the monopoles and open segments will be
heavily
damped due to the friction from the ambient plasma. The long range
magnetic field of the monopoles will be frozen into the cosmological
plasma. The tension in the open strings will shrink the segments,
bringing the monopoles and antimonopoles at the ends
together. One possibility is
that the current and the charge
in the segment would prevent the segment from
shrinking any further as happens in the case of the loop. Then the
segment would form a dumbbell and survive for a very long time.
On the other hand, if the current leaks out through the monopoles,
the segment would collapse rather quickly since the frictional
forces cannot slow down the longitudinal motion of the string
but only the transverse motion. In this scenario, the open segments
decay soon after forming and disappear. The
disappearance of open segments (and loops) would also be hastened
by the breaking up of long strings by the spontaneous nucleation
of monopole and antimonopole pairs. However, we might assume that
this process will be slow (compared to the direct collapse of a
segment) since it requires a monopole pair to nucleate by a
quantum process.

There is yet another alternative to this entire scenario which
follows from the stability analysis in Sec. 2. From Fig. 2, we
see that it is possible for the strings to be stable at
high temperatures and unstable at low temperatures. Then the
string network - the rings and dumbbells - would behave like
unstable particles with a life-time given by the time it takes
for the universe to cool down to the temperature of instability.
Unstable particles have been considered on numerous occasions
in cosmology, particularly as a means for generating additional
entropy. It is amusing that embedded strings would be natural
candidates for such unstable particles.

What may be the consequences of long-lived rings and dumbbells?
The most obvious consequence is that these objects may be the
dark matter of the universe and may still be around today. They
may be lurking in stars and in galactic halos.
On the other hand, since these objects are formed in the early
universe, there is a chance that they will come to dominate
the universe rather early (since they redshift as matter). In
this case, their cosmology might be useful to constrain particle
physics models - though, given the uncertainties, this promises
to be a difficult task. Finally, the decay of the rings and
dumbbells would produce energetic exotic particles. These decay
products might lead to interesting effects. Finally, the presence
of strings for some period of time
could lead to baryogenesis\refto{rbad}.

\beginsection{5. Conclusions}

Topological strings can have dramatic consequences
in the early universe but can occur only in
certain specially constructed particle physics models. On the other
hand, embedded strings are almost universal in their occurrence but their
consequences depend on their stability.
For the embedded string to have some
affect on cosmology, it should survive for one Hubble time at the very
least. This criterion makes it necessary to study the stability of
the electroweak Z-string at high temperatures.

We have analyzed the stability of the Z-string at high temperatures and
also taken quantum corrections to the scalar potential into account.
The analysis shows that thermal corrections tend to enhance
stability but the effect is too small to stabilize the Z-string
in the standard electroweak model with $\sin^2 \theta_W \approx 0.23$
and Higgs mass larger than $57$ GeV. Hence, we come to the conclusion
that the Z-string is unstable at all temperatures. Then, even if
a Z-string configuration is formed during the electroweak phase transition,
it will quickly decay into particles and the string will not survive
for more than a Hubble time. This means that Z-strings are probably
irrelevant for cosmology after the electroweak phase transition; their
role during the electroweak phase transition is still unclear.

We found that it is possible to construct
phenomenologically acceptable left-right models
that also admit {\it stable} embedded strings ($Z_R-$strings).
Due to its stability the $Z_R-$string may survive for a large number
of Hubble times and
may be cosmologically significant.

The cosmology of embedded strings
was discussed in Sec. 4. Here, we pointed out that embedded strings
would be produced thermally during the phase transition.
The Boltzmann suppression
of long string segments and large loops means that there is a possibility
that all the loops and segments will collapse dynamically and decay
into ordinary particles. On the other hand, the pressure from
bosonic and fermionic zero modes on the strings might prevent
this collapse and serve to stabilize the ``rings'' and ``dumb-bells''.
We considered the more interesting possibility that some of
these remnants may have survived for a few Hubble times and perhaps
even until the present epoch.

\beginsection {\it{Acknowledgements:}}

We would like to thank Ed Copeland, Rick Davis, and Mark
Hindmarsh for discussion
and the ITP, Santa Barbara where this work was partly done.
This work was supported in part by the National Science Foundation
under Grant No. PHY89-04035. RH was partially supported by DOE contract
DE-FG02-91ER40682.
SDH acknowledges support from the
National Science Foundation under grant NSF-PHY-87-14654, the state
of Texas under grant TNRLC-RGFY106 and the Harvard Society of Fellows.
The work of RW was supported in part by the NASA (NAGW-1340
at Fermilab) and by the DOE (at Chicago and Fermilab).

\references

\refis{av} A. Vilenkin, Phys. Rep. {\bf 121}, 265 (1985).

\refis{tv91} T. Vachaspati, Phys. Rev. Lett. {\bf 68}, 1977 (1992);
T. Vachaspati, ``Electroweak Strings'', TUTP-92-3.

\refis{jct} J. C. Taylor, ``{\it Gauge Theories of Weak Interactions}'',
Cambridge University Press, 1976.

\refis{hnpo} H. B. Nielsen and P. Olesen, Nucl. Phys. B{\bf{61}}, 45 (1973).

\refis{ynetc} Y. Nambu, Nucl. Phys. B{\bf 130}, 505 (1977);
N. S. Manton, Phys. Rev. D{\bf 28}, 2019 (1983); M. B. Einhorn
and R. Savit, Phys. Lett. B{\bf 77}, 295 (1978).

\refis{mjlptv} M. James,
L. Perivolaropoulos and T. Vachaspati, ``On the Stability of
Electroweak Strings'', preprint (1992).


\refis{mohapatra} See for example, R. N. Mohapatra in {\it Quarks, Leptons and
Beyond}, H. Fritzsch et al, Plenum, New York and London (1985).

\refis{buchmuller} W. Buchmuller and C. Greub, ``Right Handed Currents and
Heavy Neutrinos in high energy $ep$ and $e^{+}e^{-}$
scattering'', DESY 92-023.

\refis{seesaw} M. Gell-Mann, P. Ramond and R. Slansky in {\it Supergravity}
(North Holland, Amsterdam 1979); T. Yanagida in {\it Proceedings of the
Workshop on Unified Theory and Baryon Number of the Universe}, KEK, Japan 1979.

\refis{gelminironc} G. Gelmini and M. Roncadelli, Phys. Lett. {\bf 99B}, 411
(1981).

\refis{georgi} H. Georgi, S. Glashow and S. Nussinov, Nucl. Phys. {\bf B193},
297 (1981).

\refis{witten} E. Witten, Nucl. Phys. {\bf B249}, 557 (1985).

\refis{com1} It should be remarked that embedded string
solutions will exist in almost any particle physics model\refto{tvmb}.

\refis{sher} M. Sher, Phys. Rep. 179, 273 (1989).

\refis{dlhll} M. Dine, R. Leigh, P. Huet, A. Linde, and D.
Linde, ``Towards the Theory of the Electroweak Phase
Transition'', SLAC-PUB-5741;
G. Boyd, D. Brahm and S. Hsu, CALT-68-1795/HUTP-92-A027/EFI-92-22.

\refis{kibble} T. W. B. Kibble, J. Phys. A{\bf 9}, 1387 (1976).

\refis{ecmhnt} E. Copeland, M. Hindmarsh and N. Turok,
Phys. Rev. Lett. {\bf 58}, 1910 (1987);
E. Copeland, D. Haws, M. Hindmarsh and N. Turok,
Nucl. Phys. B{\bf 306}, 908 (1988);
D. Haws, M. Hindmarsh and N. Turok, Phys. Lett. B{\bf 209},
255 (1988).

\refis{chhhmt} C. Hill, H. Hodges and M. Turner, Phys. Rev. D{\bf
37},
263 (1988).

\refis{abtpds} A. Babul, T. Piran and D. N. Spergel, Phys. Lett.
B{\bf 202},
307 (1988).

\refis{rdps} R. L. Davis and E. P. S. Shellard, Nucl. Phys. B{\bf
323},
189 (1989);
R. L. Davis, Phys. Lett. B{\bf 207}, 404 (1988); Phys.
Lett. B{\bf 209}, 485 (1988);
Phys. Rev. D{\bf 38}, 3722 (1980).

\refis{champs} A. De Rujula et al., Nucl. Phys. B{\bf 333},
173 (1990);
S. Dimopoulos et al., Phys. Rev. D{\bf 41}, 2388 (1990);
J. Basdevant et al., Phys. Lett. B{\bf 234}, 395 (1990); A. Gould et
al., Phys. Lett. B{\bf 238}, 337 (1990).

\refis{rbad} R. Brandenberger and A. C. Davis, ``Baryogenesis from
Electroweak Strings'', BROWN-HET-862.

\refis{tvmb} T. Vachaspati and M. Barriola, ``A New Class of
Defects'', NSF-ITP-92-35.

\refis{tvavformation} T. Vachaspati and A. Vilenkin, Phys. Rev. D{\bf
30},
2036 (1985).

\refis{tvformation} T. Vachaspati, Phys. Rev. D{\bf 44}, 3723 (1991).

\refis{com1} It should be remarked that embedded string
solutions will exist in almost any particle physics
model\refto{tvmb}.

\refis{com4} The boundary conditions determine the net flux of
strings passing through the contour, that is, the algebraic
sum of the string windings.

\endreferences

\vfill
\eject

\beginsection{Figure Captions}

\item{1.} Regions of stability for various values of $P_e$.
Plotted are the boundaries between stability
and instability for (from left to right) $P_e=
0.5,0.4,0.25,0,-0.5,-1.0$.  Strings are (meta)stable for
parameters to the right of the curves.

\item{2.} Regions of stability for various temperatures for
fixed $D/{\alpha^2}=0.224$ and $E/{\alpha^2}= 0.019$.  Strings
are stable in regions below the solid lines and to the right of
the dashed line.  The dashed line corresponds to the boundary of
stability at the phase transition.

\endjnl
\end